\def\mag{\hbox{$\;.\!\!\!^m$}}
\def\to{\hbox{$\,$--$\,$}}
\def\muspc{\hskip 0.15 em}
\input psfig
\documentstyle[11pt,conf_iap]{article}
\begin{document}
\heading{COLOUR-MAGNITUDE DIAGRAMS FROM MICROLENSING PROJECTS
AND GALACTIC STRUCTURE} 
\null\vspace*{-10mm}\null
\photo{}
\null\vspace*{-10mm}\null
\author{Y.K. NG}
       {Institut d'Astrophysique de Paris, CNRS, 98bis Boulevard Arago, 
       F--75014 Paris. }

\bigskip

\begin{abstract}{\baselineskip 0.4cm 
The Colour-Magnitude Diagrams (CMDs)
from the microlensing projects provide valuable 
information, to improve our understanding of galactic structure. 
A description is given of the extinction along the line of sight.
The MACHO and OGLE
microlensing candidates (\cite{MACHO},\cite{OGLE-events})
are used to tune up
the population model and to
determine the contribution from various stellar populations
in the CMD. About 5\%, 25\%, 40\% and 30\% of the stars are 
found to be located in respectively the halo, bulge, bar and
disc of our Galaxy. The contribution from the disc is 
probably lower, if the disc density drops considerably 
between \hbox{3\to4~kpc} from the galactic centre.
}
\end{abstract}

\section{Introduction}
The HRD-GST (Hertzsprung-Russell Diagram Galactic Software Telescope)
is a new galactic structure model. It is developed 
to study the properties of the stellar populations,
which contain a wealth of information about the formation
and evolution of our Galaxy.
The analysis tool is based on the stellar population synthesis
technique. Its basis is formed by the 
latest stellar evolutionary tracks calculated by the 
Padova group \cite{Isochrones}. A smooth metallicity
coverage is obtained through interpolation between
the sets of tracks from low \hbox{(Z\muspc=\muspc0.0004)} to high 
\hbox{(Z\muspc=\muspc0.10)} metallicity. 
Figure 1 shows a schematic diagram of the HRD-GST,
see \cite{thesis} and \cite{PaperI} for additional details.
The difference between the HRD-GST and other (semi-) empirical models
is that the stars in a particular age-metallicity population
all have the same scale-height and scale-length.
For a detailed description of the results obtained thus far,
see \cite{PaperII,HHB} and \cite{DeNIS-III}\to\cite{NGC6528}.

\section{Extinction}
Any study undertaken with the HRD-GST starts
with a careful determination of the extinction. 
Figure 2a shows the extinction along the line of sight,
determined 
from the disc sequence for OGLE's Baade's Window (BW) subfield~\#3
\cite{OGLE-BW,PaperIII,Extinction}.
The blue edge of this sequence is mainly populated by the stars 
near the main sequence turn-off, while the red edge 
shows the core-H exhausted stars.

\setbox1=\vbox{\hsize=5.6cm
\null\noindent\quad
\psfig{file=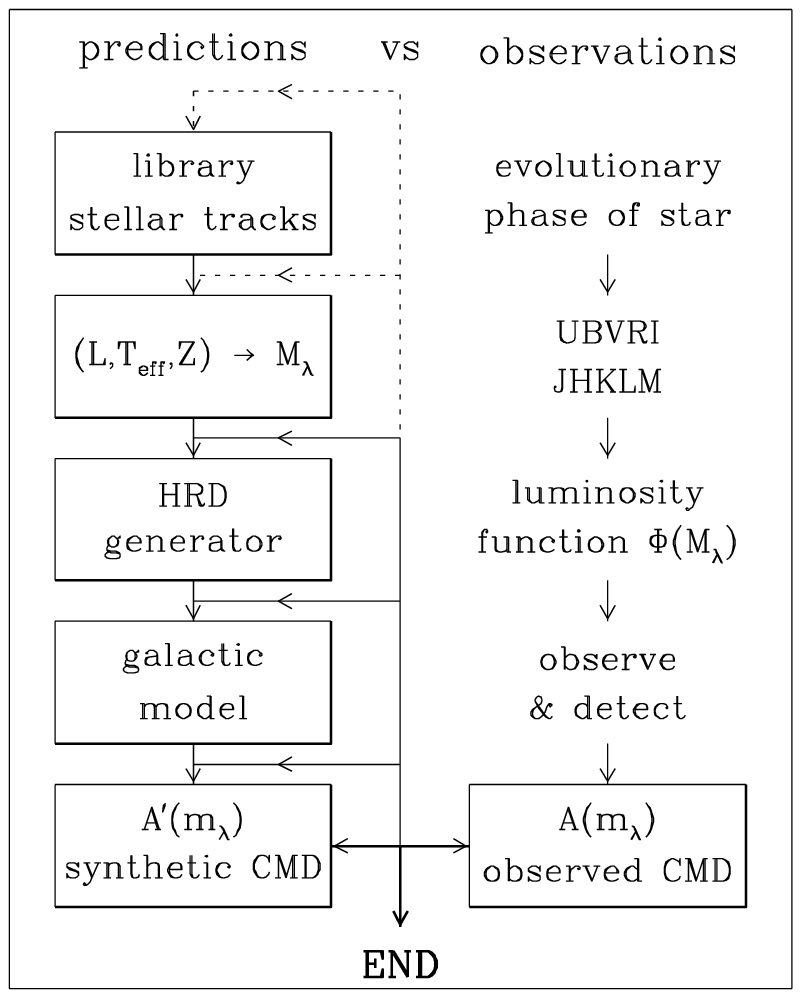,height=6.2cm,width=5.5cm}
}
\setbox2=\vbox{\hsize=10.0cm
\footnotesize
\noindent {\bf Figure 1.}\quad 
Schematic diagram of the HRD-GST. Input for the 
stellar population synthesis engine is the Padova library of
stellar evolutionary tracks 
(see \cite{PaperI,Isochrones} and references cited in those 
papers for details). 
The luminosities and effective temperature for each
synthetic star of arbitrary metallicity 
is then transformed to an absolute
magnitude in a photometric passband with the method outlined by
\cite{Bressan,Charlot}.
A synthetic HR-diagram is generated,
after specification of the stellar luminosity function 
through 
the initial mass function, the star formation rate and the
age \& metallicity range.
Synthetic stars from those diagrams are then `observed'
and `detected' with the galactic model, through a Monte-Carlo technique.
In this model the density distribution of each galactic component along the 
line of sight is specified.
This results in a synthetic CMD of the field of interest.
The synthetic CMD ought to be 
comparable with the observed CMD,
when a realistic set of input parameters is used.
If there is a marginal agreement
then check the input for each step of the HRD-GST. 
\vfill}

\centerline{\hfil\copy1\quad\quad\quad\copy2}
\bigskip\noindent
Figure 2b shows the extinction along 
the line of sight for the metal-rich
globular cluster NGC~6528 \cite{NGC6528}.
The study of the extinction along the line of sight \cite{Extinction}
complements the extinction map obtained by Stanek \cite{Stanek}.
In general, a linearly increasing extinction up to 2~kpc  
is assumed (\cite{Arp,OGLE-BW}).
Figure 2a shows, that a distance up to 4~kpc would be more appropriate.
An even better two step, linear approximation is
$$
A_V(d)=\cases{ 0.5\;d &for $d\le2$~kpc\cr
 1+{(A_V\cite{Stanek}-1)(d-2)\over3}
&for 2\muspc$<\!d$~(kpc)\muspc$\le$\muspc5\cr}\quad,
$$
where $A_V\cite{Stanek}$ refers to the value from the extinction map.
This relation is only valid for $A_V\!<$\muspc2\mag5.

\section{The Bar population}
The first indication with the HRD-GST of the bar was found in 
the study of a field in the galactic plane towards NGC~6603 
\cite{thesis,PaperII}. This study suggested 
an inclination angle $\phi=18^\circ\pm3^\circ$. 
From the red HB (horizontal branch) stars evidence was
found in the OGLE CMDs for the galactic bar \cite{OGLE-bar}. 
Its population was identified with the HRD-GST 
in OGLE's CMD of BW \cite{Bar,PaperIII}.
The age is 
\hbox{8\to9~Gyr} with an estimated uncertainty of 2~Gyr. 
This uncertainty is mainly due to the large metallicity spread,
ranging from
\hbox{Z\muspc=\muspc0.005\to0.030} among the bar stars.
The origin of this large metallicity spread is unknown. 
Studies of globular clusters \cite{NGC6440,NGC6528} 
made thus far 
do not indicate that this result is an artifact of the HRD-GST.
\par
In (V,V--I) CMDs the direction of the metallicity
gradient is almost parallel to the extinction vector. This
makes it difficult to separate one from the other. 
It is demonstrated, that a large amount of differential extinction
cannot account for the red horizontal branch (RHB) morphology 
\cite{Extinction}.
The RHB morphology is most likely a combination of differential 
extinction combined with a range in metallicities of the RHB stars.
Actually, near-infrared observations can be useful to settle this question.
Figure 3 shows a simulation for a (J,J--K) CMD with suitable photometric 
errors. It shows, that in this plane the RHB with a metallicity gradient 
is decoupled from the extinction vector.

\section{Do we have a hole in the disc?}
Indications that the disc density towards the galactic centre 
decreases significantly, while the density of the bar/bulge/halo  
continues to increase, have been found in various studies with the HRD-GST
\cite{PaperI,PaperII,PaperIII}. 
Figure~4 shows the magnitude distributions of the stars for
various the colour slices through the CMD. 
Frame~f shows that the simulated distribution for an exponential
disc near \hbox{V\muspc=\muspc19$^m$} is
about two times larger than expected from the observations. 
The results 
do not confirm a start of the decrease at $\sim$\muspc2.5~kpc distance 
\cite{OGLE-BW} from the Sun. Instead,
the decrease sets in at about 3\to4~kpc distance from
the galactic centre. \eject

\setbox1=\vbox{\hsize=8.1cm
\null\noindent
\vbox{
\psfig{file=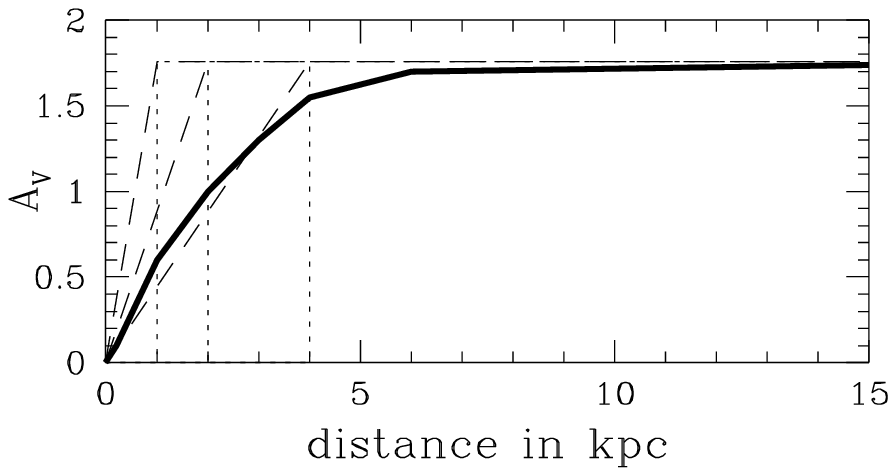,height=5.2cm,width=8.0cm}}
\hfill\break
\smallskip\noindent
\vbox{\footnotesize\noindent
{\bf Figure 2a.}\quad
The extinction along the line of sight in BW3 (thick line,\cite{PaperIII}).
Linearly increasing (long-dashed line) and step-like (dashed line) 
extinction curves for various distances are also shown. 
Details about the corresponding 
simulations can be found in \cite{Extinction}}
\hfill\break
\null\medskip\null\noindent
\vbox{\psfig{file=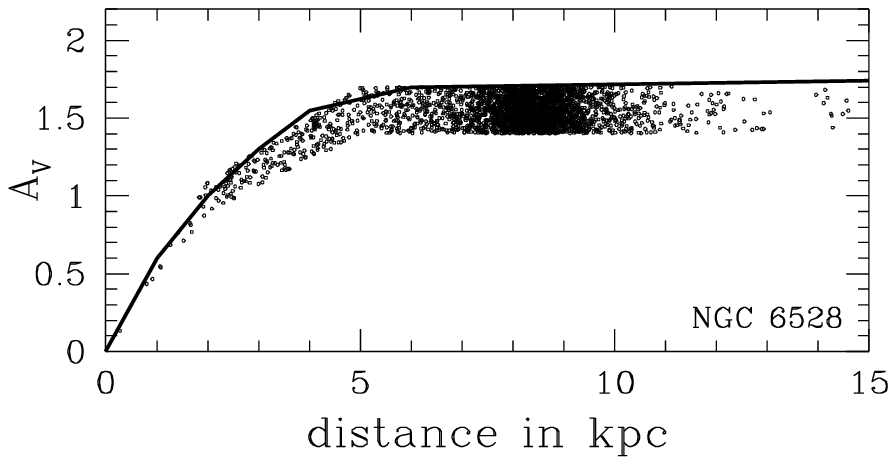,height=5.2cm,width=8.0cm}}
\hfill\break
\noindent
\vbox{\footnotesize\noindent
{\bf Figure 2b.}\quad
The extinction along the line of sight towards the metal-rich globular 
cluster NGC 6528 (dots,\cite{NGC6528}); the thick solid line shows the
extinction for BW3 \cite{PaperIII}}}

\setbox2=\vbox{\hsize=8.1cm
\null\noindent
\vbox{
\psfig{file=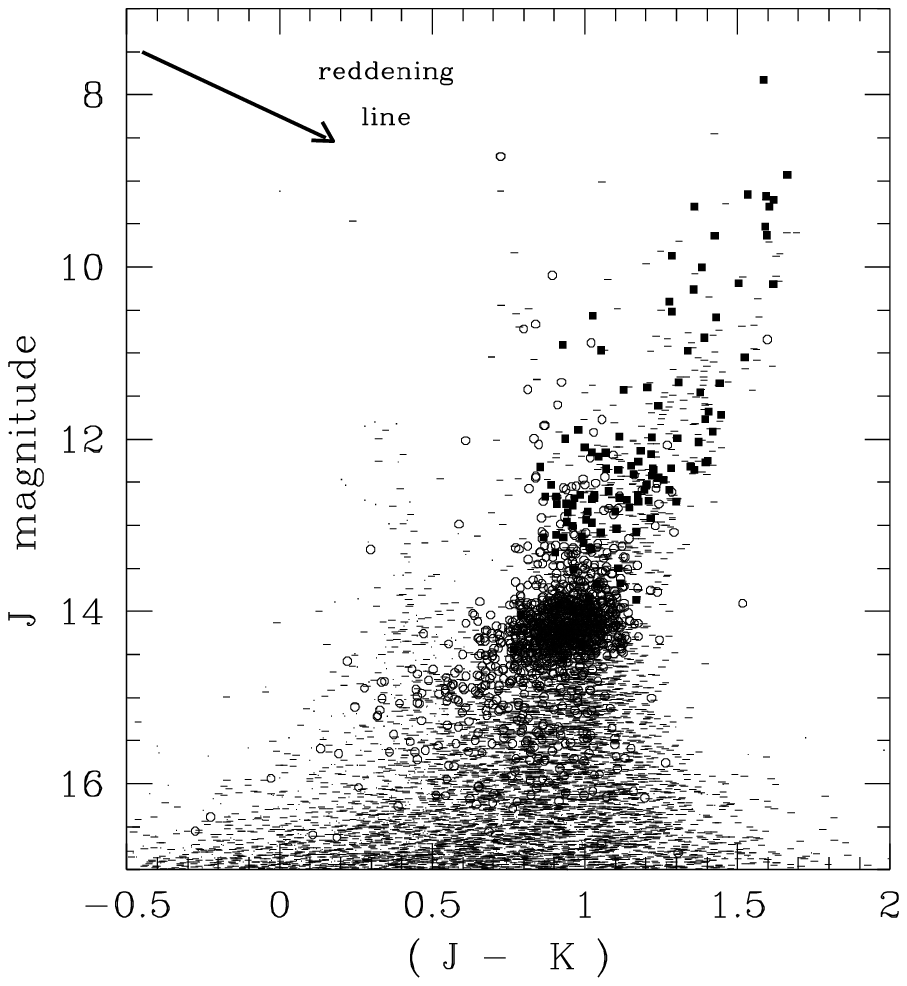,height=10.0cm,width=8.0cm}}
\hfill\break
\null\hfill\break
\smallskip\noindent
\vbox{\footnotesize\noindent
{\bf Figure 3.}\quad
Simulated (J.J--K) colour-magnitude diagram for 
Baade's Window subfield \#3.
The small dots represent the stars on the main sequence, 
the dashes the core H-exhausted or the red giant
branch stars, 
the open circles the horizontal branch stars
and the squares represent the asymptotic giant branch stars.
}}

\centerline{\copy1\hfill\copy2}
\bigskip
\noindent
Following \cite{PaperII}, this decrease can 
be described by multiplying the disc density law with
$A\exp\bigl(-\alpha(R_{hole}-r)/R_{hole}\bigr)$ for 
\hbox{$r\!\le\!R_{hole}$}, where $R_{hole}$ is the radius
at which the hole starts, $A$ is a constant calculated from the
normalization at $r\!=\!R_{hole}$, and $\alpha$ describes the power
of this decrease. This description or a power-law 
will be adopted in a more detailed
study of the decrease of the disc density.

\section{Star counts}
The microlensing candidates from the MACHO and OGLE collaborations
are used to improve the galactic population model and
to determine the relative contributions 
from the various stellar populations in the CMDs,
see \cite{PaperIII} for details about the procedure. 
This relative contribution give at the same time the relative probability
that a source is from the bar, bulge, halo or disc. 
About 5\%, 25\%, 40\% and 30\% of the stars are 
found to be located in respectively the halo, bulge, bar and
disc of our Galaxy. The probability that the lensed star 
will drop with a factor two if the decrease of the disc density 
is taken into consideration. Figure~5 shows the probability 
calculated for the bulge population, which was lower for
the first events found. This behaviour was not noticed for the
other populations. It appears that the OGLE was in the beginning
less sensitive for bulge events than MACHO. This result basically 
reflects the difference between the use of a dedicated (MACHO)
and a non-dedicated (OGLE) telescope for the project.
\par
In \cite{PaperIII} the star counts are shown for a galactic model
with and without the bar population, respectively Figs.~11.9 and 6.
From these figures one might be tempted to conclude that the model
without a bar population is in better agreement with the data than the model
with a bar and that \goodbreak

\setbox1=\vbox{\hsize=10.1cm
\null\noindent
\psfig{file=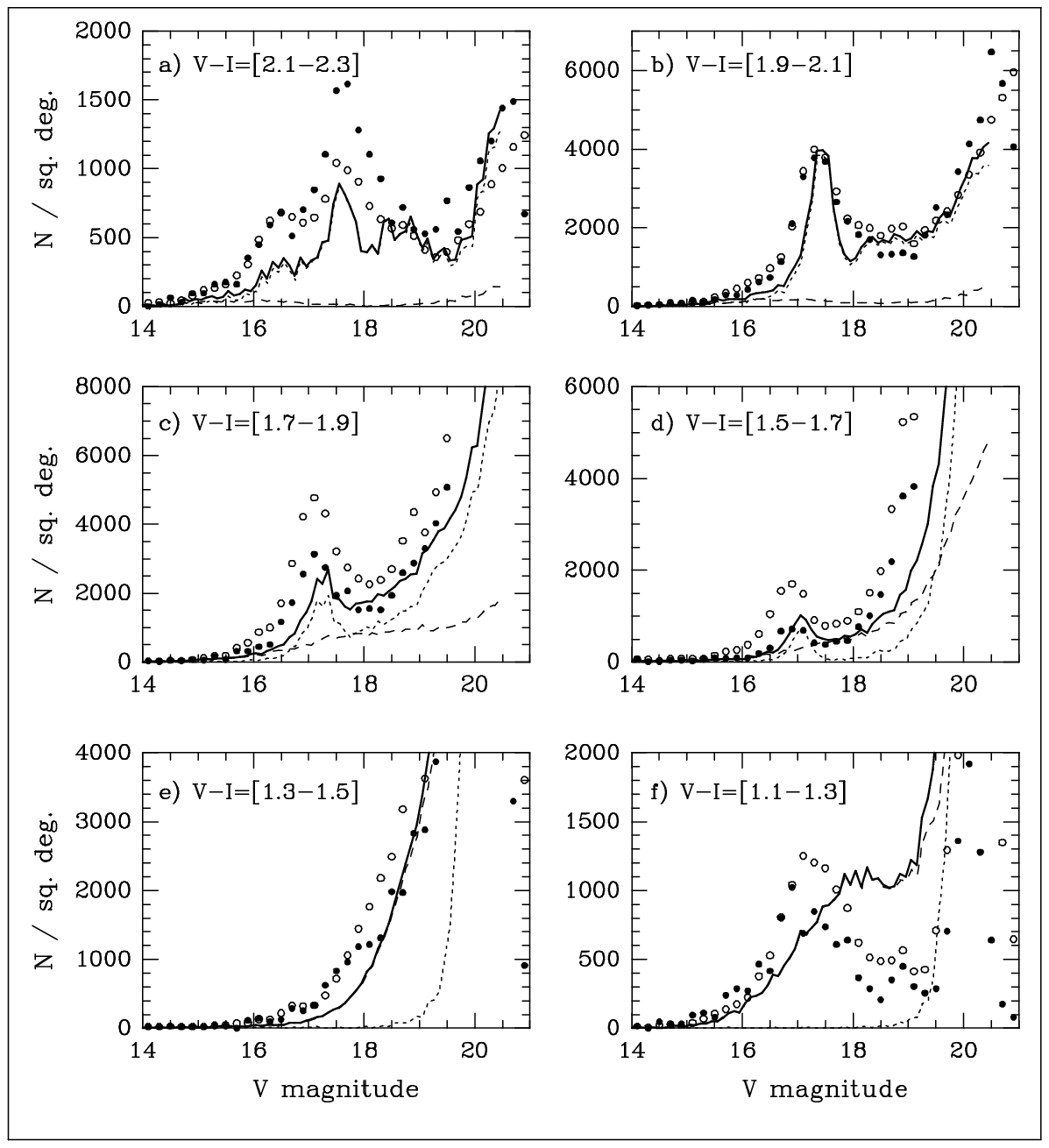,height=10.0cm,width=10.0cm}
}

\setbox2=\vbox{\hsize=6.1cm
\footnotesize
\noindent
{\bf Figure 4.}\quad
Simulated magnitude distribution (solid line) together with 
the observations for all the 9 subfields in Baade's Window
(open dots) and BW3 (filled dots). The dotted line shows the 
combined contribution from the halo/bulge and bar, while the long
dashed line gives the contribution of the disc. See \cite{PaperIII}
for details 
}

\centerline{\copy1\hfill\copy2}

\bigskip\noindent
we therefore do not have a bar. This example is 
mainly used to emphasize, that star counts alone do not mean 
anything. They are only meaningful, when shown in combination
with the colour and magnitude distributions. 
In principle, one would rely first on the direct comparison of the 
observed with the synthetic CMD. The colour and magnitude 
distributions and the star counts are only of secondary 
and tertiary importance.

\section{Age-metallicity relation for the galactic stellar populations}
The stellar populations found with the HRD-GST and used in 
the simulations are shown in figure~6. The ages and metallicities
of these populations contain information about the formation
and the metal enrichment history of the Galaxy
\cite{AMR,PaperIII}.
Apparently the star formation stopped earlier 
in the bulge (inner part) than in the halo. This is probably an
indication for replenishment with (new) barely enriched material from the 
outer parts. It implies that assumption about a closed box model might
be invalid. Secondly, the bar formed during a second, major
star formation epoch. If the bar is formed through
a merger, then the age of the bar indicates that
this event occurred about 9~Gyrs ago. Such an event would puff
up the disc. A study of the older disc populations 
could be used to investigate if they contain records
of this event. A detailed study of the CMDs towards the LMC 
is interesting, because they might contain this information. 
\par
Note, that the bulge is old. The bar population  
might have been mistakingly identified with the bulge.
An age as young as 5~Gyr does not appear to be possible.
Does this exclude an age as young as 5~Gyrs for stars in the `bulge'?
It might be an artifact, induced by the assumption of 
solar metallicity for the bulge stars. To match the lower
value of the metallicity range, one has to assume a younger age.
On the other hand,
the carbon stars in the direction of the `bulge' suggest, that
stars are present with an upper limit for the age of 5~Gyr.
To solve this dilemma, it is important to determine the nature 
of the carbon stars and verify if some of them 
are distributed inside a (flattened) spheroid.

\setbox1=\vbox{\hsize=8.1cm
\null\noindent
\psfig{file=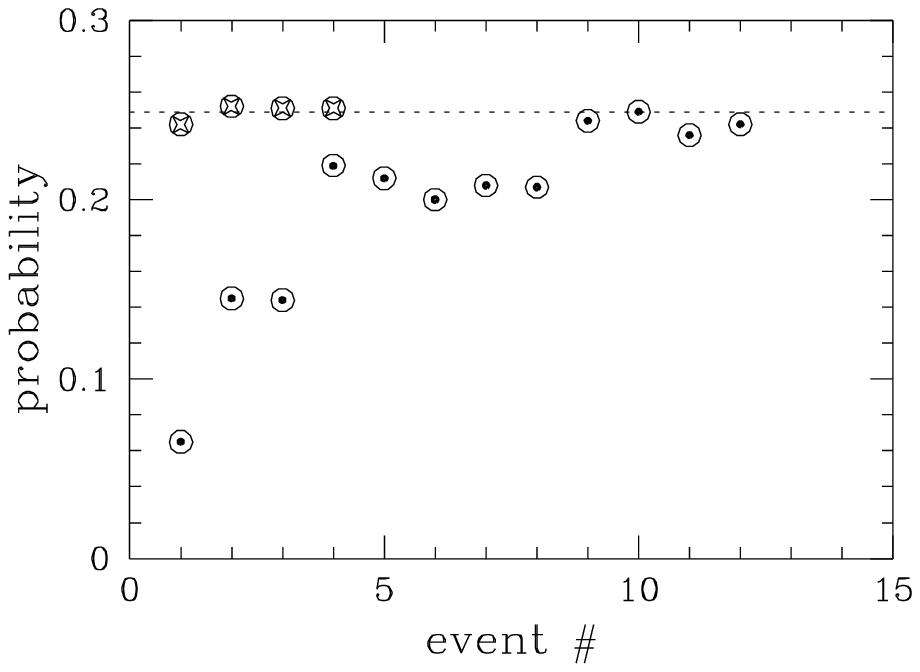,height=5.2cm,width=8.0cm}
}

\setbox2=\vbox{\hsize=8.1cm
\null\noindent
\psfig{file=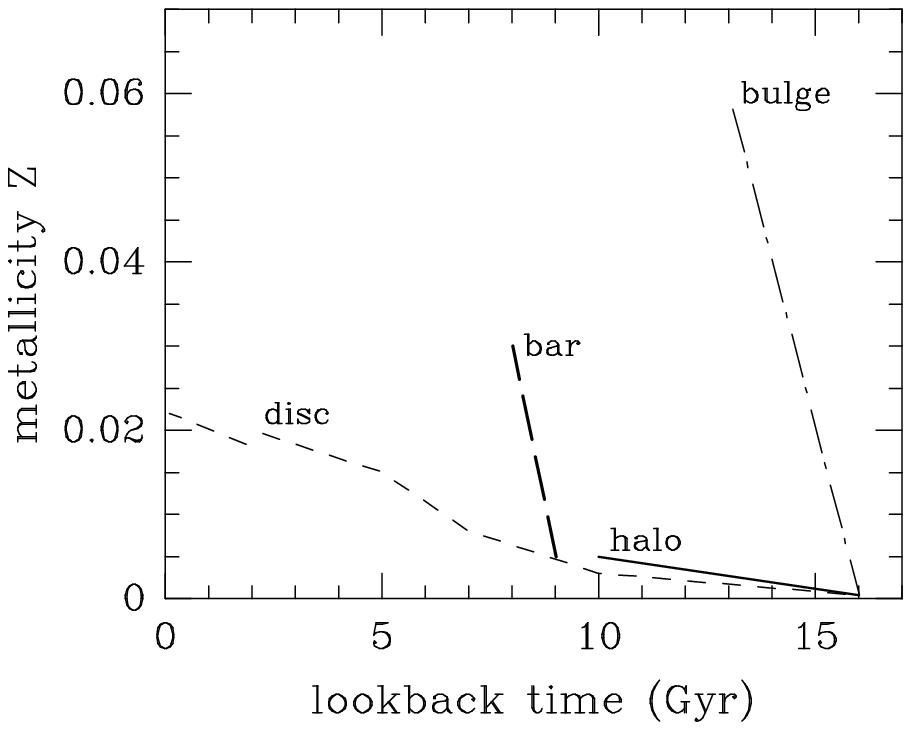,height=5.2cm,width=8.0cm}
}

\setbox3=\vbox{\hsize=8.1cm
\footnotesize
\noindent
{\bf Figure 5.}\quad
The stellar population probability derived from BW3,
that the four MACHO \cite{MACHO} 
and the twelve OGLE \cite{OGLE-events} events 
could be bulge stars
}

\setbox4=\vbox{\hsize=8.1cm
\footnotesize
\noindent
{\bf Figure 6.}\quad
The interpolated age-metallicity relation obtained thus far 
with the HRD-GST from the galactic stellar populations}

\vbox{%
\centerline{\copy1\hfill\copy2}
\smallskip
\centerline{\copy3\hfill\copy4}
}
\bigskip

\acknowledgements{Ng is supported by HCM grant CHRX-CT94-0627
from the EC.}

\vfill
\end{document}